\documentclass[conference]{IEEEtran}
\IEEEoverridecommandlockouts
\usepackage{cite}
\usepackage{amsmath,amssymb,amsfonts}
\usepackage{algorithmic}
\usepackage{graphicx}
\usepackage{textcomp}
\usepackage{xcolor}
\def\BibTeX{{\rm B\kern-.05em{\sc i\kern-.025em b}\kern-.08em
    T\kern-.1667em\lower.7ex\hbox{E}\kern-.125emX}}
\begin{document}

\title{Innovations in Nanotechnology: A Comprehensive Review of Applications Beyond Space Exploration}

\author{\IEEEauthorblockN{1\textsuperscript{st} Batool Zehra Ladha}
\IEEEauthorblockA{\textit{Department of Electrical Engineering} \\
\textit{Habib Uniersity}\\
Karachi, Pakistan \\
bl07621@st.habib.edu.pk}
\and
\IEEEauthorblockN{2\textsuperscript{nd} Syed Muhammad Muslim Hussain}
\IEEEauthorblockA{\textit{Department of Electrical Engineering} \\
\textit{Habib University}\\
Karachi, Pakistan \\
sh07730@st.habib.edu.pk}
\and
\IEEEauthorblockN{3\textsuperscript{rd} Muhammad Hasan Khan}
\IEEEauthorblockA{\textit{Department of Electrical Engineering} \\
\textit{Habib Uniersity}\\
Karachi, Pakistan \\
mk07712@st.habib.edu.pk}

}
\maketitle

\begin{abstract}
Nanotechnology has emerged as a transformative force across multiple industries, enhancing materials, improving instrumentation precision, and developing intelligent systems. This review explores various nanotechnology applications, including advancements in materials science, healthcare, energy storage, environmental monitoring, and robotics. Nanomaterials, such as carbon nanotubes and graphene, offer significant improvements in fields like energy generation and medicine, while nanosensors revolutionize environmental and industrial monitoring. Micro and nano robots provide automation solutions across industries. By expanding beyond space exploration, this review highlights the far-reaching potential of nanotechnology to reshape industries through interdisciplinary collaboration and innovation.
\end{abstract}
\begin{IEEEkeywords}
 Nanotechnology, Materials Science, Healthcare, Energy Storage, Environmental Monitoring, Nanorobotics, Nanosensors, Nanomaterials
\end{IEEEkeywords}
\section{Introduction}
Nanotechnology has become a pivotal tool in numerous scientific and industrial fields due to its ability to engineer materials at the atomic and molecular levels, offering transformative properties across diverse industries \cite{lane2011nanobots}. While originally investigated for its potential in space exploration, nanotechnology's applications extend far beyond, playing a critical role in fields such as healthcare, energy storage, environmental monitoring, and manufacturing \cite{benaroya2002nanotechnology} . The term “nanotechnology” describes a range of technologies performed on a nanometer scale, where materials exhibit unique physical, chemical, and biological properties compared to their bulk counterparts \cite{lozano2015nanoengineered} \cite{janson2003micro}.

The enhanced properties of nanomaterials, such as their lightweight, strength, and durability, provide opportunities to create materials that outperform conventional ones \cite{scalia2020nanomaterials}. In healthcare, nanomaterials are being used to develop targeted drug delivery systems, diagnostic tools, and nanomedicine. Energy sectors benefit from nanotechnology through advancements in battery technology, solar cells, and fuel cells, which rely on nanostructured materials to improve energy storage and efficiency \cite{arnold2005nanostructured} \cite{gohardani2014potential}. In environmental applications, nanosensors enable precise monitoring of pollutants, while in manufacturing, nanotechnology has driven innovations in nanomanufacturing processes that produce smaller, more efficient devices \cite{loomis2005thermal}\cite{globus2000molecular}.

Increased demand for miniaturized systems and micro-products has significantly driven the growth of research in micro- and nano-manufacturing \cite{dalla2022carbon}. Nanomanufacturing has enabled the creation of highly complex parts crucial to several industries, such as medical devices and electronics [11]. Nanotechnology's ability to create materials with tailored properties has resulted in advancements like super-strong, lightweight materials that enhance energy efficiency in transportation and manufacturing systems \cite{dalla2022carbon}. Furthermore, it has enabled the development of ultra-small, highly capable robotic systems, impacting industries like automation and healthcare \cite{delfini2017cvd}. In addition to transforming industries on Earth, nanotechnology continues to hold promise for space applications by reducing the size, weight, and cost of spacecraft components, though its greatest potential now lies in other sectors \cite{santoli1999hyper}.

Nanomaterials, engineered at the atomic and molecular level, offer unique properties that are revolutionizing a wide range of applications. From energy systems and propulsion technologies to nanomedicine and environmental sensors, nanotechnology can potentially change the landscape of multiple industries [15]. At the nanoscale, materials exhibit properties distinct from those observed in their bulk forms, such as increased strength, higher surface area, and altered chemical reactivity \cite{Arepalli_Moloney_2015}. This literature review explores the diverse applications of nanotechnology, focusing on its use in medicine, energy storage, environmental monitoring, manufacturing, and robotics, providing insight into its transformative potential in various fields.

\section{Nanotechnology in Space Exploration}
The literature review for this research can be divided into the following four applications of nanotechnology in space exploration:

\begin{itemize}
    \item Spacecraft Architecture
    \item Nanosensors \& Instrumentation
    \item Micro/Nano Robotics
\end{itemize}

\subsection{Spacecraft Architecture}
Nanotechnology has transformed space exploration by facilitating the creation of small vehicles or micro-crafts designed for space missions.\cite{lane2011nanobots}. These developments are particularly evident in the application of nanobots from hydrobots to enter Europa's ice surface, demonstrating the transformative impact of nanotechnology in space exploration. \cite{benaroya2002nanotechnology}. The utilization of nanomaterials in microspace structures not only enhances functionality but also reduces the cost of space missions, as volume and mass directly correlate with expenses \cite{lozano2015nanoengineered,janson2003micro}.

Nanotechnologies offer significant advantages across various space segments, including reduced vehicle mass, improved functionality, and enhanced propulsion performance \cite{scalia2020nanomaterials}. NASA's experimentation with nanotechnology extends to the development of nano-structured heat shields, which offer superior thermal insulation compared to conventional materials \cite{arnold2005nanostructured}. Carbon nanotubes, with their exceptional mechanical, thermal, and electrical properties, hold significant potential for future space vehicles \cite{gohardani2014potential}.



The heat shields used by NASA are made of fully dense carbon phenolic (CP). Just adding 5\% of nano-fiber by volume does not change the properties of the material but it shows an increase in thermal conductivity from 0.55 W/m\textdegree K to 500 W/m\textdegree K. The thickness of CP to sustain 250\textdegree C is 4.65 cm. However, a decrease in 30\textdegree C is observed when nano-fibers are added. This reduction demonstrates the passive heat pipe effect, where heat moves from the hot stagnation point to the cooler downstream area of the heat shield. It is estimated that this effect leads to an overall mass savings of 5-10\% in the heat shield \cite{arnold2005nanostructured}.

One area of significant advancement is the utilization of carbon nanotube (CNT) based composites in spacecraft structures. \cite{rast} discusses the development of CNT-based composites for the Juno spacecraft, highlighting the progress made in CNT technology over the past decade by Lockheed Martin Space Systems for integration into composite components of spacecraft structures. CNTs have the potential to significantly reduce weight while maintaining or even improving structural strength \cite{song2013structural} and explore the integration of various nanocarbon materials, including multi-walled carbon nanotubes (MWNT), carbon nano-onions (CNO), and graphene nanoplatelets (GnP), into composite structures \cite{dalla2022carbon}. 





Carbon nanostructures can be utilized for protecting carbon fiber reinforced polymer (CFRP) composites used in Low Earth Orbit (LEO) from atomic oxygen (AO) erosion \cite{delfini2017cvd}. The nanostructures were grown using chemical vapor deposition (CVD). A crucial aspect of the investigation involved methane chemical vapor deposition (CVD) on catalyzed carbon fiber-based substrates to cultivate the growth of carbon nanostructures. The morphology and growth characteristics of these nanostructures were thoroughly examined, encompassing parameters such as yield, purity, homogeneity, and coating uniformity \cite{delfini2017cvd}. By characterizing carbon nanostructures-reinforced carbon composites through on-ground atomic oxygen simulation facilities, the aim was to optimize the process of producing carbon-multiscale advanced composites \cite{SANTOLI1999117}.

The effectiveness of the coating process was assessed through AO ground tests. The results revealed that deposition of high-yield carbon nanofilaments significantly improved performance in terms of total mass loss and AO erosion rate compared to less-ordered carbon deposits \cite{delfini2017cvd}. This highlights the potential of polymer nanocomposites in spacecraft applications, where their unique properties – attributed to the small size and high surface area of the incorporated fillers – offer distinct advantages \cite{bhat2021review}.


Electric propulsion systems are crucial for spacecraft navigation and deep-space exploration \cite{arepalli2015engineered}. Nanomaterials are proposed to have multiple uses in the thrusters such as surface healing, ultra nanocrystalline diamonds or graphene possibly as a substitute for the ceramic wall, and graphene nanotubes for channel wear resistance \cite{patelelectric}. Self-healing, crucial for next-gen thrusters, involves surface repair of the chassis. Selective patching material deposition or electric field-induced reconstruction are proven methods for surface damage repair \cite{kumar2014electric}. 

\cite{levchenko2018recent} proposes the concept of an adaptive thruster that leverages the unique properties of nanomaterials like cluster-grown carbon nanotubes and graphene. This concept offers the potential for efficient in-flight adjustments and paves the way for more sophisticated spacecraft applications. Notably, small satellites, or CubeSats, benefit immensely from miniaturized electric propulsion systems enabled by nanomaterials, reducing launch costs and expanding space exploration capabilities \cite{koo2007polymer}.

NanoFET can also be used in making the electric propulsion system for spacecraft. In nanoFET, precisely sized and charged nanoparticles are suspended in a low vapor pressure liquid. This approach offers great flexibility in adjusting the charge-to-mass ratio to fine-tune propulsion performance \cite{gallimorescalable}. The use of nanoFETs provides a highly integrated system, geometrically scalable system, longer operational lifetime, and better controllability.

\begin{figure}
    \centering
    \includegraphics[width=0.8\linewidth]{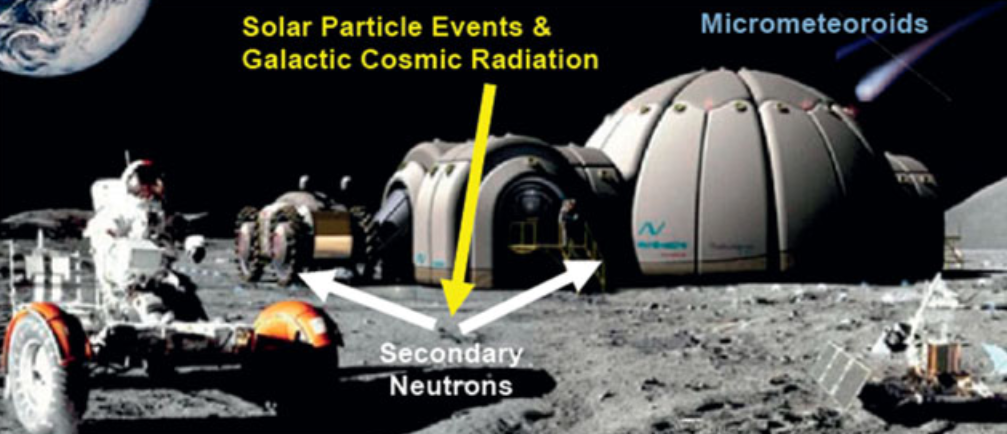}
    \caption{Space environmental hazards: space radiation and micrometeoroids.}
    \label{space-hazards}
\end{figure}


The low Earth orbit (LEO) and very low Earth orbit (VLEO) environments pose significant challenges to spacecraft due to factors like space debris, extreme temperatures, radiation, and atomic oxygen erosion \cite{nano13111763}. Graphene \cite{zhao2021exploring} and carbon nanotubes \cite{atar2015atomic} exhibit exceptional strength, thermal conductivity, and resistance to atomic oxygen erosion, making them ideal candidates for spacecraft protection.

The space environment contains major hazards to space travel, among which are space radiation and micrometeoroids, as depicted in Figure \ref{space-hazards}. \cite{Thibeault_Kang_Sauti_Park_Fay_King_2015} provide a comprehensive examination of the utilization of nanomaterials for radiation shielding in space environments, aimed at safeguarding individuals from the diverse hazards of space radiation. Nanomaterials can protect astronauts and spacecraft electronics from the detrimental effects of space radiation, including galactic cosmic radiation (GCR), solar particle events (SPEs), and secondary neutrons produced by GCR and SPE interactions with matter \cite{babu2023nanotechnology}.

\cite{cheraghi2021boron} identifies Boron Nitride Nanotubes (BNNTs) as a promising candidate for space radiation shielding. The primary source of space radiations are GCR and SPEs, but when they interact with the spacecraft structures, they produce neutron radiation. Boron isotope \(^{10}B\) has the highest neutron absorption cross section, a feature that is also higher in nitrogen when compared with carbon. The neutron absorption cross section in pure \(^{10}B\) is 3890 barns, which is much higher compared to 760 barns in naturally occurring Boron. The reaction of thermal neutron capture is given as:

\begin{equation}
    ^{10}B + n \rightarrow ^{4}He(1.47 \, \text{MeV}) + ^{7}Li(0.84 \, \text{MeV}) + \gamma(0.84 \, \text{MeV}).
\end{equation}

As a result, the thermal neutron capture property of BN is mainly displayed by boron, which results in considerable protection against neutron radiations with high \(^{10}B\) boron content in compounds. This makes Boron Nitride Nano Tubes a viable option for protection from radiation in space environments.



The interplay between thermal and electrical energy at small scales can strongly influence the functional behavior of many types of devices such as direct energy conversion elements, heat sinks, and field-effect transistors \cite{kumar2015nanotechnology}. Nanomaterials hold immense promise for advancing energy generation \cite{chen2012nanomaterials}, storage \cite{zhang2013nanomaterials}, and transmission \cite{elcock2007potential} in space applications. \cite{Arepalli_Moloney_2015} explore these advancements, focusing on improvements in solar cell efficiency and lifespan using silicon and emerging materials like graphene and perovskites. They highlight the evolution of solar cell technology, from traditional silicon cells to flexible solar arrays, demonstrating the continuous pursuit of efficient and adaptable power generation solutions for spacecraft \cite{fu2018flexible}. Furthermore, innovations in energy storage with high energy density and long-life lithium-sulfur batteries offer exciting possibilities for future space missions \cite{xu2018carbon}. Finally, lightweight conducting wires made from CNTs and wireless energy transmission via microwaves are transforming energy transmission systems in space, paving the way for a future with efficient and reliable power distribution \cite{stolojan2018manufacturing}.

\subsection{Nanosensors \& Instrumentation}

Within the wider context of nanotechnology's impact on space exploration, nanosensors and instrumentation appear as critical components with the potential to revolutionize our understanding of the universe \cite{sirbook}. This section discusses the early phases of advancement, important hurdles, and future goals and aims in this field. 

A major area where we expect to see good returns on investment is producing a large number of materials in space exploration \cite{10.1117/12.850486}. With the help of nanotechnology, we can create lightweight and durable materials. These materials are crucial for designing spaceships that are efficient in fuel usage and successful in missions \cite{10.1117/12.850486}. It's important to consider how these materials behave in space, including their physical and electromagnetic properties. \cite{bilhaut2009assessment}. Nanoscale sensors provide unprecedented precision in data collecting, allowing for full analysis of celestial body features and ambient circumstances, altering our understanding of the universe with much higher operating temperatures than current devices and sensors \cite{meyyappan2015nanoelectronics}. 

Integration procedures facilitate the seamless introduction of nanosensors and equipment into spacecraft systems, streamlining operations and increasing reliability in the harsh environment of space \cite{1597934}. Nanotechnology enables the creation of tiny biological research laboratories onboard spacecraft, allowing for experiments that lead to advances in astrobiology and life sciences research \cite{cinti2023reviewing}. Next-generation vehicle sensor systems powered by nanotechnology allow spacecraft to travel through space more precisely and safely, reducing risks and ensuring mission success in unpredictable space conditions \cite{1597934}. Rapid technological infusion facilitates the timely integration of cutting-edge nanosensor and instrumentation breakthroughs, resulting in continual innovation and success in space exploration efforts \cite{sirbook}

\begin{figure}
    \centering
    \includegraphics[width=0.8\linewidth]{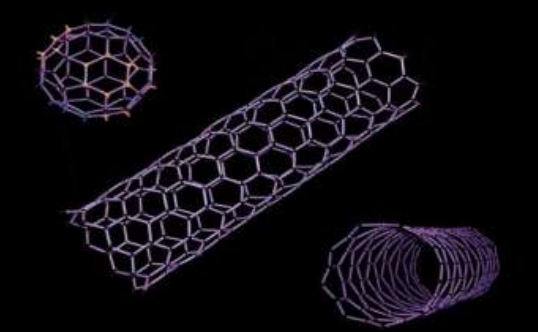}
    \caption{Carbon Nanotubes in Space Exploration}
    \label{Nano-sensors}
\end{figure}

Despite the progress made, space exploration still encounters many difficult challenges. Production and refinement provide substantial challenges, In space exploration, carbon nanotubes shown in the Figure \ref{Nano-sensors} serve as nanosensors, but their use requires refinement due to the limited production of nanomaterials such as carbon, metal, and oxide nanotubes \cite{Samareh_2017, article}. Manipulating and controlling nanoscale materials necessitates nanoscale handling procedures, complicating macroscale processes \cite{gleiter2000nanostructured}. Reliable nanoscale deposition tools are critical, demanding the development of novel technologies like inks, masks, and nanoindentations \cite{sirbook}. Standardized interfaces and unique technologies are required to bridge the gap between the macroscale and nanoscale worlds for nano-micro-macro integration, which may involve wireless communication and networked infrastructures \cite{gleiter2000nanostructured}. Establishing efficient and dependable structures becomes critical, demanding fault-tolerant designs to handle anomalies inherent in nanoscale production while also improving radiation hardness through a variety of capabilities \cite{meyyappan2015nanoelectronics}. Large sensor networks necessitate self-calibrating capabilities and efficient networking technology, and data fusion issues similar to those encountered in enormous data warehouses necessitate interdisciplinary solutions that go beyond traditional aerospace specialties \cite{scalia2020nanomaterials}. Despite these obstacles, overcoming them promises to unleash the full potential of nanosensors and instruments in propelling space research to new heights of discovery and comprehension \cite{meyyappan2015nanoelectronics}.

To address these difficult challenges, numerous significant goals have been established, to push the boundaries of space exploration through the strategic use of nanotechnology \cite{meyyappan2015nanoelectronics}. Continued demonstration of mission-enabling technologies is a primary objective, with a focus on creating new materials like band-gap-tailored semiconductors and increasing radiation shielding via electron-phenomenon decoupling. Improved passive components, including fatigue-resistant wire and high-density capacitors, are critical to increasing spaceship reliability and performance \cite{klonicki2023instrumentation}. Improving device performance in terms of sensitivity\cite{sirbook}, selectivity, weight, power, and volume is a primary priority that drives sensor technology research. Novel techniques, including fault-tolerant designs and distributed sensor networks, are being developed to reduce the impact of radiation while improving durability, reliability, and predictive maintenance capabilities. Sensor self-calibration and fabrication process control using new approaches such as fluidic self-assembly and electrostatic-assisted assembly are also being considered to improve sensor performance and reliability \cite{meyyappan2015nanoelectronics}. These efforts aim to lay the foundation for future progress in space exploration, pushing the boundaries of what we can achieve beyond the next ten years.

Micro-electro-mechanical systems (MEMS) are used in space because they're small, light, and require little power. Examples show their past and future use in space missions. The next step is nano-electro-mechanical systems (NEMS), which bring new challenges, like dealing with previously ignored forces. An example of a nano-mechanical memory is discussed. While nanosystems aren't always better than current small systems, they'll be useful in making smart networks for space missions \cite{bilhaut2009assessment,tang1998micromechanical}.

The integration of nanosensors and instruments in space research opens up possibilities, but it isn't without challenges. Challenges including a limited supply of nanomaterials and the difficulty of managing tiny components necessitate novel solutions \cite{sirbook}. Collaborative efforts are required to improve production methods, provide reliable tools, and build strong architectures. Despite these challenges, strategic goals centered on technology demonstrations, passive component improvements, and approaches such as fault-tolerant designs and sensor self-calibration represent a potential road forward \cite{sirbook}. Overcoming these problems will lead space exploration into a new era of discovery and thus, influencing future generations.

\subsection{Micro/Nano Robotics}
Micro and nanorobotics, driven by advancements in nanotechnology, will revolutionize space exploration by enabling the development of highly miniaturized, intelligent, and autonomous robotic systems. This literature review explores the potential of micro/nanorobotics in space applications, drawing insights from both a research paper by Papadopoulos et al. \cite{Papadopoulos2013} and a report by the National Nanotechnology Initiative Workshop \cite{sirbook}.

Robotic devices able to perform tasks at the nanoscale (i.e., scale of a nanometer) are called “NanoRobots.” The field of nanorobotics studies the design, manufacturing, programming, and control of nanorobotic systems \cite{mavroidis2012nanorobotics}. \cite{Papadopoulos2013} discusses the specific contributions of nanotechnology to space robotics, highlighting its role in miniaturization and the development of nanorobots. These robots, measuring between 0.1 and 10 micrometers, can be engineered using biological microorganisms or magnetic nanocapsules and guided by advanced imaging systems. Their applications range from precise manipulation at the cellular level to tasks like Mars extravehicular activity (EVA) repair, showcasing the transformative potential of nanotechnology in space exploration. The National Nanotechnology Initiative Workshop report emphasizes the broader impact of miniaturization using nanomaterials and microelectronics. It highlights advancements in power generation, allowing for miniaturized robots with reduced power requirements. This not only reduces costs but also paves the way for deploying large collectives of small robots for diverse space missions \cite{sirbook}.

\begin{figure}
    \centering
    \includegraphics[width=0.75\linewidth]{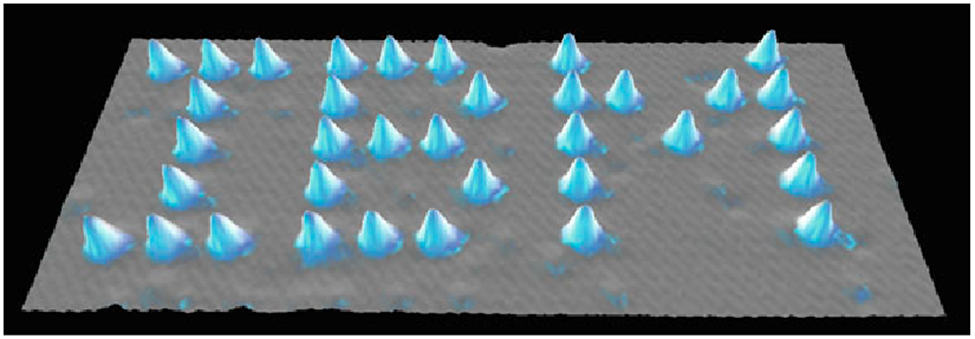}
    \caption{IBM written using atom nanomanipulation}
    \label{ibmnanol}
\end{figure}

The first nanorobotic systems were not “nano” at all. Instead, they were large manipulator-like structures that had the capability of nanopositioning. The first example of atom nanomanipulation using an surface tunnelling microscope (STM) was performed by Eigler and Schweizer in \cite{eigler1990positioning} where they nanopositioned 35 Zenon atoms to write the name of IBM, their employer, as shown in figure \ref{ibmnanol}.
Dong, Arai, and Fukuda in 2001 where they demonstrated successful operation of a 10-DOF NRM system \cite{dong20013}. They then developed a 16-DOF nanorobotic manipulator (NRM) system \cite{fukuda2003assembly} equipped with a SEM for real-time imaging of the manipulation task and a nanofabrication system based on electron-beam-induced deposition \cite{fukuda2005nanorobotic}. These nano manipulators can help in miniaturizing the robotic systems for space, an important motivation for the systematic miniaturization of robotic systems is that launch vehicles have tight constraints concerning the payload’s mass and volume characteristics and therefore successful miniaturization directly results in an improved, more compact and less expensive system \cite{papadopoulos2012miniaturization}. Space systems in general and space robotics in particular have the following sub-systems: Power, Propulsion, Structure, Attitude and Orbital Control (AOCS), On-Board Data Handling (OBDH), Locomotion, Guidance, Navigation and Control (GNC), Communication, Thermal, Manipulators and End – Effectors \cite{sellers2005understanding}.

One way to remotely control the nanorobotics systems in space would be to use magnetically guided nanorobotics systems. By manually or automatically adjusting the position or orientation of a magnet, a translatory or rotational movement of MagRobots can be triggered. Direct utilization of portable magnet provides an easy-to-operate way to drive the motion of MagRobots by simply adjusting the position and orientation of a magnet \cite{zhou2021magnetically}. \cite{menon2007space} discusses a nature-inspired climbing nanorobot that has gecko-like micro structured hairs to explore different types of terrains on different planets. 

\begin{figure}
    \centering
    \includegraphics[width=0.75\linewidth]{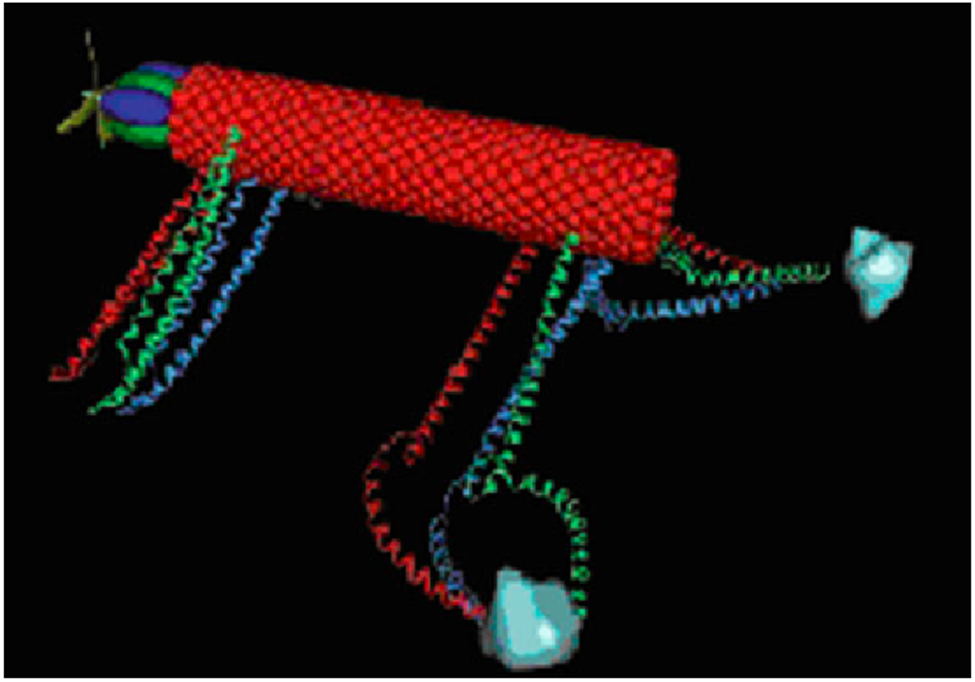}
    \caption{A vision of a bio nanorobotic organism}
    \label{biobot}
\end{figure}
Bionanorobotics systems can also be employed in space exploration. The term bio nanorobotics, which was first introduced in 2003 \cite{dubey2003viral,dubey2004dynamics}, denotes all nanorobotic systems that include nanocomponents that are based on biological elements such as proteins and DNA as shown in figure \ref{biobot}. These components perform their preprogrammed biological function in response to the specific physiochemical stimuli but in an artificial setting. In this way, proteins and DNA could act as motors, mechanical joints, transmission elements, or sensors. If all these different components were assembled together in the proper proportion and orientation they would form nanorobotic devices with multiple degrees of freedom, able to apply forces and manipulate objects in the nanoscale world \cite{mavroidis2012nanorobotics}. Networked TerraXplorers (NTXp) is a concept in which a network of channels containing the bio nanorobots having enhanced sensing and signaling capabilities \cite{mavroidis2005space}. Another bionanosystem is the All Terrain Bio-nano Gears, depicted in Figure 12, which acts as an added protective layer for the human body. It can detect harmful environments, such as radiation or chemicals, well before they pose a serious threat. Acting as both an early warning system and a means of administering treatment, it assists in healing and prevents damage to astronauts \cite{mavroidis2005space}.
\begin{figure}
    \centering
    \includegraphics[width=0.75\linewidth]{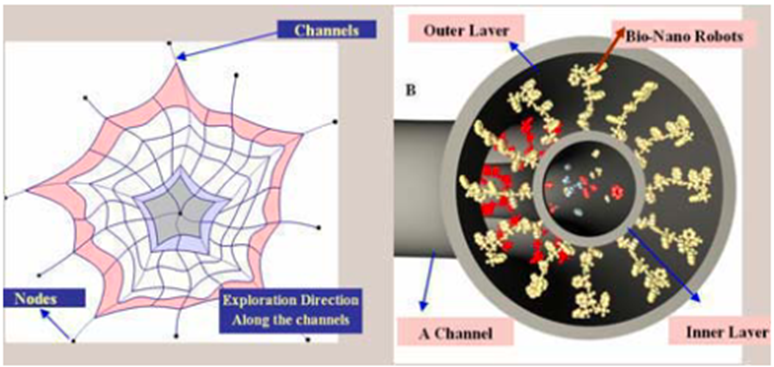}
    \caption{A Networked TerraXplorers (NTXp)}
    \label{NTRp}
\end{figure}

\section{Conclusion}
In conclusion, this review highlights the broad impact of nanotechnology across various fields, from healthcare to energy storage and industrial automation. Nanomaterials, nanosensors, and nanorobotics have proven transformative in these sectors, providing solutions for more efficient, precise, and sustainable technologies. As interdisciplinary collaboration continues and nanotechnology evolves, its potential to revolutionize industries and solve pressing global challenges grows ever stronger.

\bibliographystyle{IEEETran}
\bibliography{main}


\end{document}